# Twin Mechanical Metamaterials

Wenwang Wu[1,2*#], Seok Kim[1,3*], Ali Ramazani[1*], Young Tae Cho[3], Daining Fang[4#]

1, Department of Mechanical Engineering, Massachusetts Institute of Technology, Cambridge, MA 02139, USA.

2, School of Naval Architecture, Ocean and Civil Engineering, Shanghai Jiao Tong University, Shanghai, 200240, China.

3, Department of Mechanical Engineering, Changwon National University, Changwon, South Korea.

4, Institute of Advanced Structures & Technologies, Beijing Institute of Technology, Beijing, 100084, China.

\* equally contribution

\# corresponding authors

Wenwang Wu       (wuwenwang@sjtu.edu.cn)

Daining Fang     (fangdn@pku.edu.cn)

**Abstract**

By mimicking the geometrical relation of nano-twin crystals, we propose novel architected twin mechanical metamaterials (TMMs), which can impede local shearing band formation under external loading, thus avoiding global catastrophic failure. The effects of twin-space and twin angle on the mechanical performance of TMMs were also explored, such as: energy absorption, strength, and crack propagation resistance. The results showed that the twin topology design can not only significantly improve the energy-absorption efficiency but also remarkably improve the crack-propagation resistances of stretching-dominant mechanical metamaterials. We also studied the effect of twin-space and twin angle on the tensile strength of TMMs. This study is the first to report on the inverse Hall–Petch effect of TMMs. Our findings open an avenue for the design and fabrication of advanced materials with exceptionally tuneable mechanical properties.

**Keywords:** mechanical metamaterials; energy absorption; Hall-Petch effect; twin; mechanical properties; crack propagation resistances.

**Introduction**

Natural and artificially manufactured lightweight structured materials and structures with robust strength, stiffness, energy-absorption efficiency, and toughness, have been actively used over the past thousands of years. In recent decades, various types of advanced lightweight structured alloys, such as dual-phase steel [1–4], twin-induced plasticity steel [5, 6], and transformation-induced plasticity steel [6, 7], have been developed for industrial applications. The well-established Hall–Petch strengthening and inverse Hall–Petch effects have been observed in different types of nano-crystalline materials, and various types of theoretical models have been proposed for exploring the underlying strengthening and softening mechanisms [8-10]. However, there are intrinsic conflicts between strength, ductility, and toughness; thus, multi-scale multi-mechanisms, such as twin, graded, and hierarchical structural strategies, based on micro-structural designs of advanced alloys with extraordinary comprehensive mechanical performances are of critical importance. In recent decades, twin microstructures have demonstrated a promising potential for improving the comprehensive mechanical performances of advanced alloys, such as: strength [11, 12], toughness [11, 12], or fatigue resistance [13]. In general, twin configurations are categorized into three types based on the different twin spatial relations of elements [14]. By mimicking the graded structures of biomaterials, graded nano-crystalline microstructures with spatially evolving, constituent geometrical features can reduce the catastrophic failures of materials, demonstrating the best combination of strength and toughness [15, 16]. According to the graded features of structural design strategies, former studies have proposed structures with novel grain-size, twin-thickness, lamellar-thickness, columnar-size, platelet, and layered gradients [17]. Normally, twin structures are beneficial for toughness and graded structures are beneficial for strain hardening and strength improvement [18–20]. Based on the synergistic design of twin and graded structural features, extraordinary mechanical benefits can be produced. (a) twin structure results in a stable and history-independent cyclic response governed by a new type of dislocation called the correlated necklace dislocation [13, 20]. (b) gradient nano-twin structure results in additional strengthening that breaks the rule of mixture composite theory, governed by a new type of dislocation mechanism called bundles of concentrated dislocations [19, 20]. Moreover, nano-crystalline materials can develop hierarchical twinned microstructures, which can provide additional blocks of dislocation movement for strengthening the polycrystalline metal, leading to a higher strength compared with the individual nano-twinned metals [21–24].

In recent years, innovative mechanical metamaterials with complex artificial microstructures have been proposed for generating remarkable stiffness, strength, improved impact energy absorption, and enhanced fracture toughness [25–30]. Owing to the progress of 3D printing techniques at the micro- and nano-scales, various types of novel architected mechanical metamaterials can be designed and manufactured; for example, micro-lattice metamaterials with extraordinary damage tolerances [31–37], hierarchical lattice mechanical metamaterials with ultra-strength and stiffness [38, 39], shellular metamaterials with ultra-strength and stiffness [40], plate-lattice metamaterials with extraordinary stiffness [41–43], and multi-walled mechanical metamaterials for energy absorption [44]. Recently, polycrystalline-microstructure-inspired lattice mechanical metamaterials with shear band suppression effects were developed for damage-tolerance enhancement [45]. In this study, by mimicking the geometrical relations of polycrystalline microstructures and biomaterials with internal twin topological features, as shown in **Fig. 1 (a)–(c)**, we designed and demonstrated novel twin mechanical metamaterials (TMMs), including ordinary, graded, and hierarchical twinning architectures via projection microstereolithography (PμSL)-based 3D printing techniques (**Fig. S1**). In-situ compression experiments were performed to study the deformation mechanisms and energy-absorption performances of TMMs. Similarly, twin mechanical metamaterials (TMMs) with gradient twin angle and twin space can also be designed and fabricated, as shown in **Fig. S2(a)-(b)**. We also investigated the effect of topological parameters of TMMs on the mechanical behaviours of ultimate tensile strength and crack-propagation resistances.

**Energy-absorption performances of TMMs**

The energy-absorption efficiency of the impact-attenuating material is expressed as the ratio of the area under the compressive stress–strain curve (energy absorbed) divided by the maximum stress achieved up to a given strain level. In recent decades, various types of lightweight materials and structures have been developed for energy-absorption applications, for example, porous metals, honeycombs, lattice structural materials, or mechanical metamaterials [25, 26, 44]. Foams or bending dominant lattice impact energy attenuators deform via bending of cell walls or constituent beams. The specific strength and stiffness are much lower, and the peak stress is slightly higher than the mean plateau stress. This exhibits a relatively remarkable energy-absorption efficiency and low mean plateau stress level, which is close to the ideal response for a standard flat-on-flat impact scenario, and the energy per unit volume absorbed is equal to the area under the strain-stress curve and is always less than the ideal absorbed energy for a flat-on-flat impact. Stretching-dominant lattice impact energy attenuators deform via buckling and exhibit post-peak softening features in their strain-stress response, thus exhibiting relatively high specific stiffness and strength, and demonstrating a high mean plateau stress level and relatively low energy-absorption efficiency. In this study, seven samples (labelled **A, B, C, D, E, F, G**) were designed and fabricated to study the compression mechanical performances of ordinary, graded, hierarchical TMMs and the effects of meta-predicates on energy-absorption indicators. The details of these samples are described in **Table S1**. First, we performed comparisons between ordinary TMMs and periodically architected face-centred cubic (FCC) mechanical metamaterials, and the strain–stress curves of samples **A, B,** and **C** are shown in **Fig. 1(j)**. As shown, for sample **A**, which is a standard periodically architected FCC mechanical metamaterial, the stress first increases until the peak value and then drops remarkably until it reaches the plateau stress level, resulting in relatively low energy-absorption efficiency. The compression-based deformation continues with slight increase in the plateau stress until the densification strain level is reached, at which point the compression stress increases sharply with respect to compression strain. In contrast, the peak stress level of TMMs, i.e., samples **B** and **C**, drops slightly until it reaches the mean plateau stress. Thus, their energy-absorption efficiency is much higher than that of the standard FCC metamaterials. TMMs based on stretching-dominant FCC unit cells can maintain the high mean plateau stress of the stretching-dominant lattice with a reduced peak stress value, thus overcoming the intrinsic conflicts between high peak stress and low energy-absorption efficiency of the lattice. Moreover, twin interfaces can impede the formation of a global shearing band and restrain local structural failures within each twin layer, thus avoiding global catastrophic failures. The local deformation and failure process of TMMs (sample **B**) during the compression process are shown in **Fig. S3**.

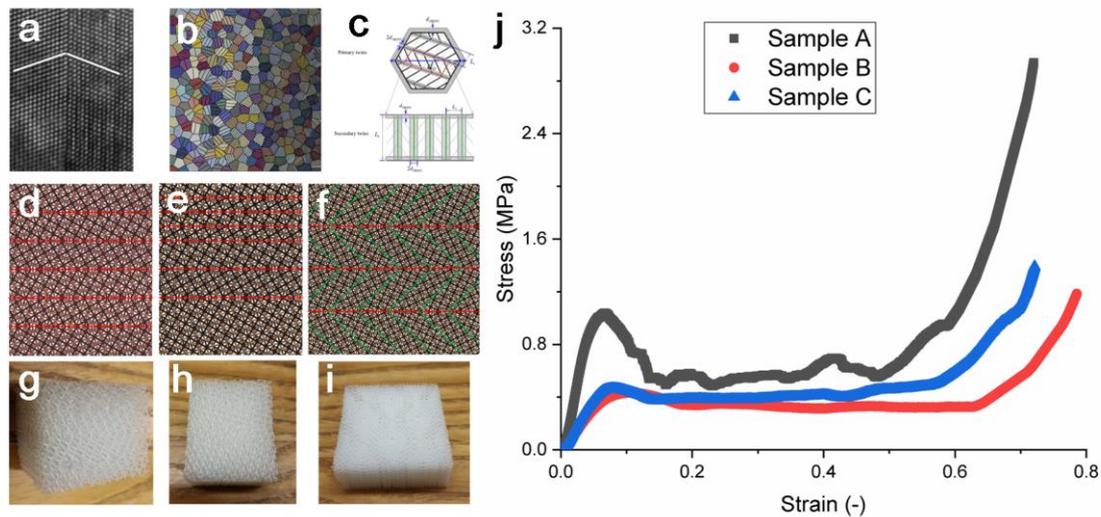

**Fig. 1: Twin interface in advanced alloys and deformation behaviour.** (a) Ordinary twin interface [22, 24]; (b) graded twin polycrystalline microstructures; and (c) hierarchical twin microstructures consisting of different levels. Novel TMMs: (d) mechanical design and (g) fabrication of ordinary twin structures with equally spaced twin-interface design; (e) mechanical design and (h) fabrication of TMMs with graded twin interface spaces; and (f) mechanical design and (i) fabrication of hierarchical TMMs consisting of 1st- and 2nd-order twin interfaces. (j) compression strain–stress curves of standard FCC periodical architected mechanical metamaterials and TMMs with even twin-spaces.

In-situ compression tests were performed for studying the deformation process of graded samples **D** and **E**, and comparisons were performed between the strain–stress curves of the graded (samples **D** and **E**) and ordinary (samples **B** and **C**) TMMs for understanding the mechanical benefits of the graded twin design, as shown in **Fig. 2(a)**. The graded design can amplify the mean plateau stress level, thus improving the energy-absorption efficiency of ordinary TMMs with equal twin spacing. As shown in the local deformation process and failure features of graded TMMs (sample **D**) in **Fig. S4**. As shown in **Fig. 2(b)**, compression tests were performed to study the energy absorption performances of hierarchical TMMs consisting of primary and secondary twin topological features (**Fig. S5**). The plateau stress of hierarchical TMMs (samples **F** and **G**) is lower than ordinary TMMs (samples **B** and **C**). Moreover, the energy-absorption efficiency of hierarchical TMMs is remarkably improved than those of ordinary and graded TMMs. Meanwhile, the energy-absorption performances, such as peak stress, mean stress, and energy-absorption efficiencies, of different types of TMMs were compared. The standard FCC periodic architected sample produces a higher peak stress than all TMMs, and the differences in the mean plateau stress of all samples are relatively small, as shown in **Fig. 2(c)**. Regarding the energy-absorption efficiency, all TMMs demonstrate increased efficiency compared to standard FCC periodically architected mechanical metamaterials without the twin design, as shown in **Fig. 2(d)**. The energy-absorption efficiency of standard FCC mechanical metamaterials without the twin design (sample **A**) is $\varphi = 0.59$ and those of ordinary TMMs (samples **B** and **C**) with identical twin-spaces are $\varphi = 0.77$ and $\varphi = 0.86$, respectively. Further, the energy-absorption efficiency of the graded TMMs (samples **D** and **E**) are $\varphi = 0.86$ and $\varphi = 0.90$, respectively. Afterwards, this efficiency of TMMs can be further improved by implementing hierarchical design (samples **F** and **G**), achieving $\varphi = 0.89$ and $\varphi = 0.96$, respectively. Finally, compression tests of TMMs with both graded twin spaces and twin angles are show in **Fig. S2(c)**, exhibiting remarkable energy absorption efficiency $\varphi = 0.94$.

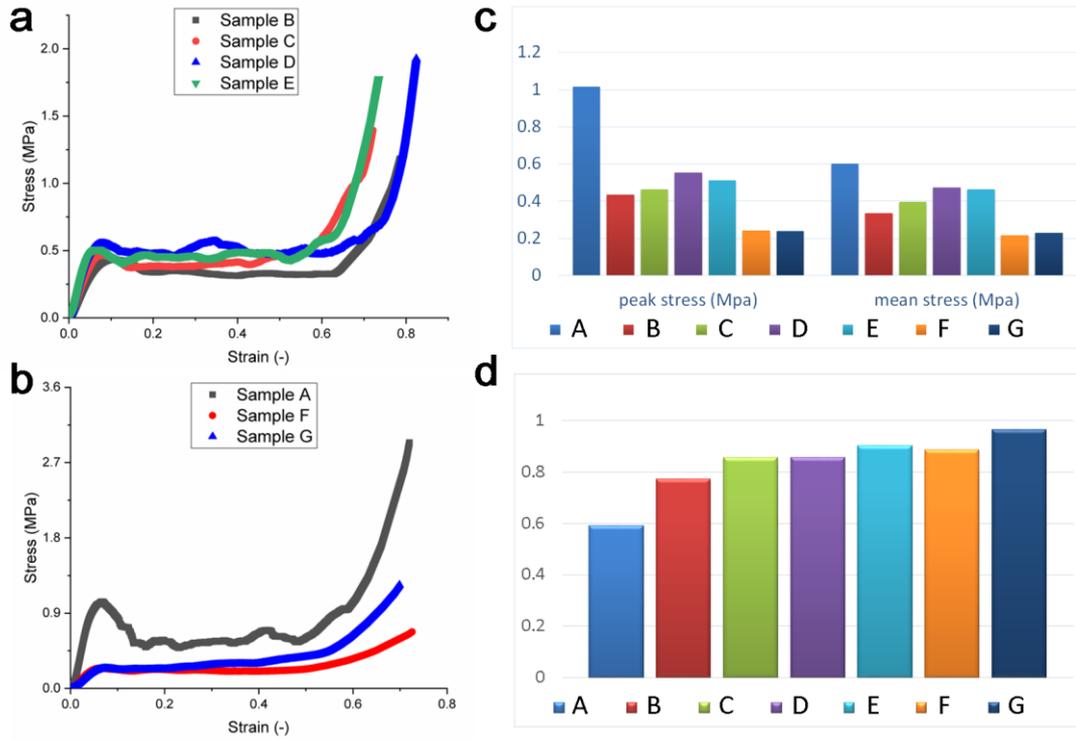

**Fig. 2: Compression test of TMMs.** (a) Comparison of the strain–stress curves of ordinary TMMs and graded TMMs. (b) Compression energy-absorption strain–stress curves of hierarchical TMMs. Comparison of the energy absorption performances indicators of ordinary twin, graded twin, and hierarchical TMMs. (c) Graphical representation of peak stress, mean plateau stress, and (d) energy-absorption efficiency.

**Tensile strength of TMMs**

To understand the effects of twin-space width on the ultimate tensile strength of TMMs, we designed test samples based on the FCC unit cell (cell size, $s$ = 2.0 mm; beam diameter, $d$ = 0.25 mm), where the twin-space width is: $t$ = 2, 3, 4, 5, 6, 7, and 8 mm. For each twin-space design, the twin angles are given as $\theta$ = 9°, 18°, 27°, and 36°. The length of the sample test section was $L$=24 mm, and width of the samples was $W$ =16, 15, 16, 15, 18, 14, and 16 mm, respectively. For each sample type, five identical samples were fabricated and repetitive tensile tests were performed to harvest the average tensile strength. As shown in **Figs. 3(b)–(g)**, the ultimate strength increases with an increase in the twin-space width from 2 to 4 mm, exhibiting an inverse Hall–Petch effect. With the further increase in the twin-space width from 4 to 8 mm, the ultimate strength decreases, exhibiting the Hall–Petch effect. **Figure 3(h)** shows the deformation process and failure features of a typical sample. The co-existences of Hall-Petch effect and inverse Hall-Petch effect can be explained with the competition between twin interface defects density and shearing band formation within each twin layer. When twin space is smaller than threshold value for generating peak strength, twin interface defects density is quite high and is responsible for ultimate strength reduction, thus the demonstrating inverse Hall-Petch effect; With the increase of twin space larger than threshold value for generating peak strength, shearing band formation effects will become dominant, demonstrating Hall-Petch effect.

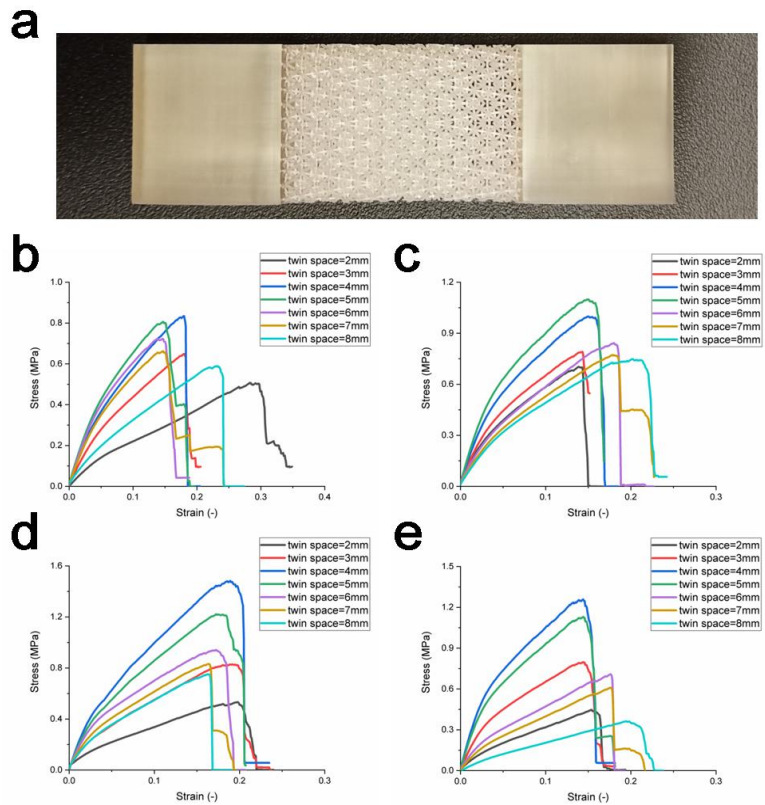

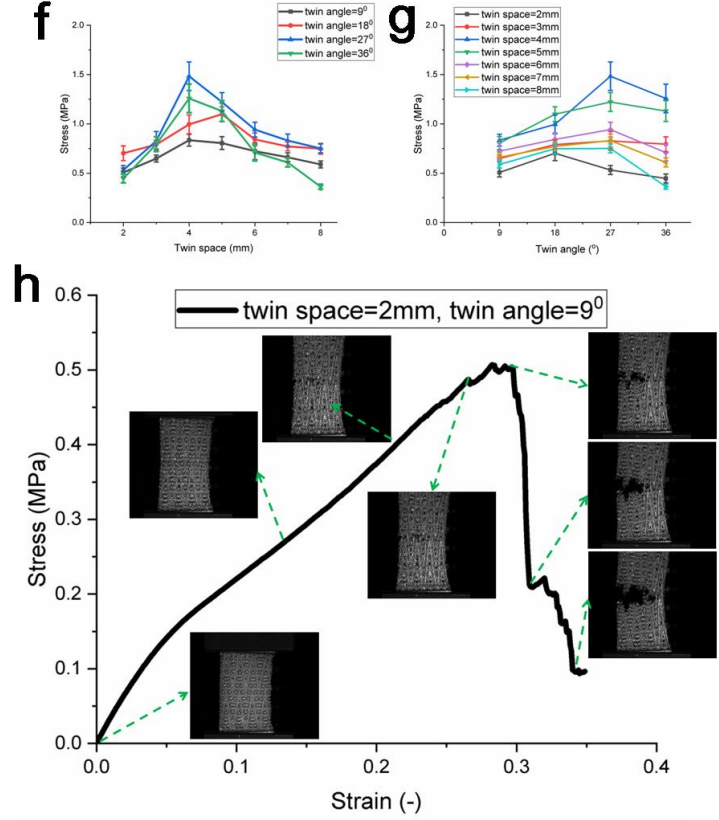

**Fig. 3: Tensile test of TMMs.** (a) As-fabricated typical sample for tensile test. Typical strain–stress curves of TMMs for studying effects of twin-space and twin angles of (b) $\theta$=9°, (c) $\theta$=18°, (d) $\theta$=27°, and (e) $\theta$=36°. Relation between twin topology and ultimate strength: (f) effects of twin-space width on the ultimate strength; (g) effects of twin angle on ultimate strength; (h) typical strain–stress curve and microstructure evolution process during the tensile test procedure (twin-space width **t**=2mm, and twin angle $\theta = 9°$).

**Crack-fracture toughness of TMMs**

Three-point bending experiments were performed to investigate the crack-fracture toughness benefits of twinning the mechanical metamaterials. The FCC-type lattice unit cell (cell size, $s$ = 2.0 mm; beam diameter, $d$ = 0.25 mm) was selected for constructing one standard FCC sample and six twinning FCC 3-point bending samples, where twin-space width, $t$= 2 and 4 mm and twin angle, $\theta$ = 9°, 18°, and 27°. In total, seven samples with the same global geometrical layout were designed and fabricated, as shown in **Fig. 4(a)**. The length, height, and thickness of the as-fabricated samples were 56 mm, 16 mm, and 8 mm, respectively. A notch with length $L_c$ = 4.3 mm was formed on one side of the printed sample, located in the middle of the sample. Next, three-point bending tests were performed to test the crack-propagation resistances. The resultant displacement–force curves are shown in **Fig. 4(b)**, and the typical crack-propagation within the TMM sample during the 3-point bending test is shown in **Fig. 4(c)**. The ratios of ultimate strength and energy absorption of twin mechanical metamaterials and periodically architected face-centred cubic (FCC) mechanical metamaterials are shown in **Figs. 4(d)** and **4(e)**, it can be seen that compared to standard FCC-based lattice structures without a twin topology, the twin design can improve the ultimate fracture strength and energy-absorption capability, thus improving the crack-propagation resistances.

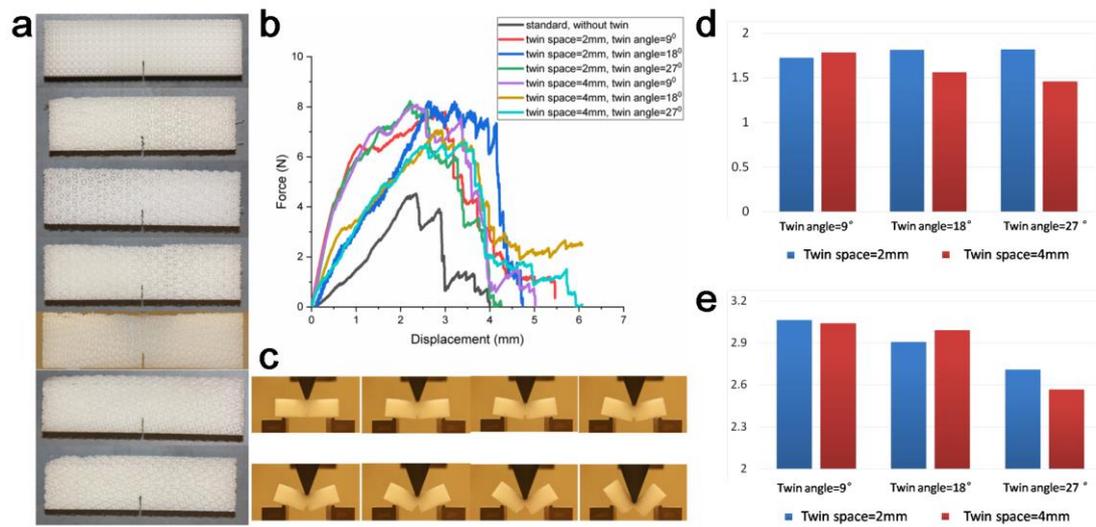

**Fig. 4: Three-point bending test for TMMs**. (a) As-fabricated standard FCC lattice sample without a twin design, and as-fabricated twin FCC mechanical metamaterials with twin angles of $\theta$= 9°, 18° and 27° and twin-space widths of $t$ = 2 and 4 mm. (b) Displacement–force curves of as-designed and as-fabricated twin FCC mechanical metamaterials. (c) Typical deformation and failure process of TMMs with the twin angle of $\theta$= 27° and twin-space width $t$ = 4 mm at displacement value 0.6, 1.2, 1.8, 2.4, 3.0, 3.6 and 4.2 mm. (d) Comparison between the 3-point bending ultimate strengths of the TMMs and standard ordinary periodic FCC reference sample. (e) Comparison between the 3-point bending energy absorption of the TMMs and standard ordinary periodic FCC reference samples.

**Crack resistances of graded TMMs**

Based on the graded twin crystalline features [47], graded TMMs with identical twin angles of $\theta$ = 30° and an increasing twin-spaces were fabricated based on the FCC unit cell (cell size, $s$=2.4mm; beam diameter, $d$ = 0.3 mm), where the widths of the three twin layers are $t_1$ = 2.4 mm, $t_2$ = 4.8 mm, and $t_3$ = 7.2 mm. The TMMs consist of four $t_1$ layers, two $t_2$ layers, and one $t_3$ layer. The length of the experimental sample test section was $L$ = 24mm, the width of the sample was: $W$ = 19.2 mm, and the thickness of the sample was 4.8 mm. The crack length was measured as $L_c$ = 2.4 mm, and it was situated on one side of the tensile sample. As shown in **Fig. 5(a)**, when the initial artificial pre-existing crack occurs on the coarse twin-layer side, the crack-propagation resistance from the coarse twin layer to the fine twin layer is much stronger than the crack propagation resistance from the fine to coarse twin sides. The energy-absorption ability of the graded sample with an artificial pre-existing crack on the coarse twin-layer side during the crack propagation was approximately twice that of the sample with an artificial pre-existing crack on the fine twin-layer side. Details of the crack-propagation process from the coarse twin layer to the fine twin layer and vice versa are shown in **Figs. 5(b)** and **5(c)**, respectively.

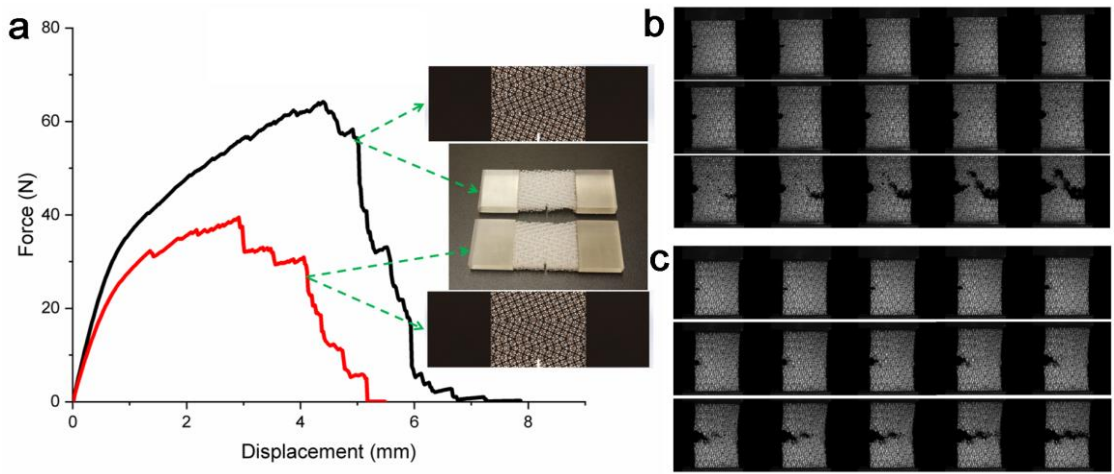

**Fig. 5: Uniaxial tensile strain–stress curves of TMMs containing one-sided cracks.** (a) Tensile strain–stress curve of graded twin lattice, with crack on one side; (b) crack propagates from coarse twin to fine twin direction, with graded twin-space width of $t$ =7.2, 4.8 and 2.4 mm (twin angle, $\theta$ = 30°; crack length $L_c$ = 2.4 mm). Local deformation features at the strain levels of $\varepsilon$ = 0.0, 0.0188, 0.0396, 0.0603, 0.1018, 0.1252, 0.1472, 0.1680, 0.1894, 0.2114, 0.2211, 0.2315, 0.2412, 0.2524, and 0.2742, respectively. (c) Crack propagates from fine twin to coarse twin direction, with graded twin-space of $t$ = 2.4, 4.8, and 7.2 mm (twin angle, $\theta$ = 30°; crack length $L_c$ = 2.4 mm). Local deformation features at the strain levels of $\varepsilon$ = 0.0, 0.0047, 0.0259, 0.0472, 0.0691, 0.0911, 0.1130, 0.1356, 0.1576, 0.1809, 0.1935, 0.2035, 0.2152, 0.2261, and 0.2410, respectively.

**Relation of crack-propagation resistances with crack features and twin topology**

To understand the effects of twin-space width and twin angle on the crack-propagation resistances of TMMs, two sample series with a crack size of $L_c$ = 4 mm were designed and manufactured based on the FCC unit cell (cell size, $s$ = 2.0 mm; beam diameter, $d$ = 0.25 mm). For the first series, the twin-space widths were set as $t$ = 2, 4, 6, and 8 mm, with identical twin angles set as $\theta$ = 18°. The corresponding lengths of sample test sections were obtained as $L$ = 20, 20, 18, and 16 mm; and the cross-section widths of the samples were $W$ = 20, 20, 18, and 16 mm, respectively. The identical sample thickness was 4 mm. For the second series, the twin-space widths were set as $t$ = 4 mm, with the twin angles of $\theta$ = 9°, 18°, 27°, and 36°. The corresponding length of sample test sections was obtained as $L$ = 20 mm, and the cross-section width of the samples was $W$ = 20 mm. The identical sample thickness was 4 mm. As shown in **Fig. 6(a)**, when changing the twin-space from 2 to 8 mm with increasing steps of 2 mm, the crack resistance for the sample with twin space of 4 mm was the highest; the ultimate fracture strength first increases and then decreases with increasing twin-space. As shown in **Fig. 6(b)**, when the twin angle is changed from $\theta$ = 9° to 36° at an increasing angular step of 9°, the crack resistance for the twin angle of $\theta$ = 18° is the highest, where the ultimate fracture strength first increases and then decreases with increasing twin angle until $\theta$ = 36°. The microstructure evolution process and failure features of a typical sample are shown in **Fig. S6** for the sample with a pre-existing crack of $L_c$= 4 mm, situated in the middle, with the twin-space width of $t$= 4 mm and twin angle of $\theta$ = 18°. To understand the effects of crack orientation on the crack-propagation resistances of TMMs, a set of samples with crack sizes of $L_c$= 4 mm and different spatial orientations with angles $\beta$ = 0°, 22.5°, 45°, and 67.5° were designed and fabricated based on the FCC unit cell (cell size, $s$ = 2.0 mm; beam diameter, $d$ = 0.25 mm). The twin-space widths were set as $t$ = 4 mm, with identical twin angles of $\theta$ = 18°. The corresponding samples width and corresponding length of sample test sections were $W$ = 20 mm and $L$ = 20 mm, respectively. The sample thickness was constant at 4 mm. As shown in **Fig. 6(c)**, when changing the spatial angle of the crack from $\beta$ = 0° to 67.5° at angular steps of 22.5°, the ultimate fracture strength first decreases, and the crack resistance for the twin angle of $\beta$ = 22.5° is the weakest. With the further increase in the spatial angle, the crack-propagation resistance increases until $\beta$ = 67.5°. The typical microstructure-evolution process and failure features are shown in **Fig. S7** for the sample with a pre-existing crack of $L_c$=4mm and spatial orientation angles of $\beta$ = 0°, 22.5°, 45,° and 67.5° situated in the middle of the sample; the twin-space width is $t$= 4 mm and the twin angle is $\theta$ = 18°. To understand the effects of crack size on the crack-propagation resistances of TMMs, a set of samples with crack sizes of $L_c$= 2, 4, 6, and 10 mm were designed and manufactured based on the FCC unit cell (cell size, $s$ = 2.0 mm; beam diameter, $d$ = 0.25 mm). The twin-space widths were set at $t$ = 4 mm with identical twin angles of $\theta$ = 18°. The corresponding samples cross-section width is $W$ = 20 mm, with sample corresponding length of samples test sections $L$ = 20 mm and identical sample thickness 4 mm. As shown in **Fig. 6(d)**, when changing the crack size from 2, 4, 6, and 10 mm, the crack resistances for crack size $L_c$ = 2 mm are the highest, and the ultimate fracture strength decreases with increasing crack size. The microstructure evolution process and failure features are shown in **Fig. S8** for the samples with pre-existing cracks with increasing crack sizes in the middle of the samples; twin-space width $t$ = 4 mm and twin angle $\theta$ = 18°.

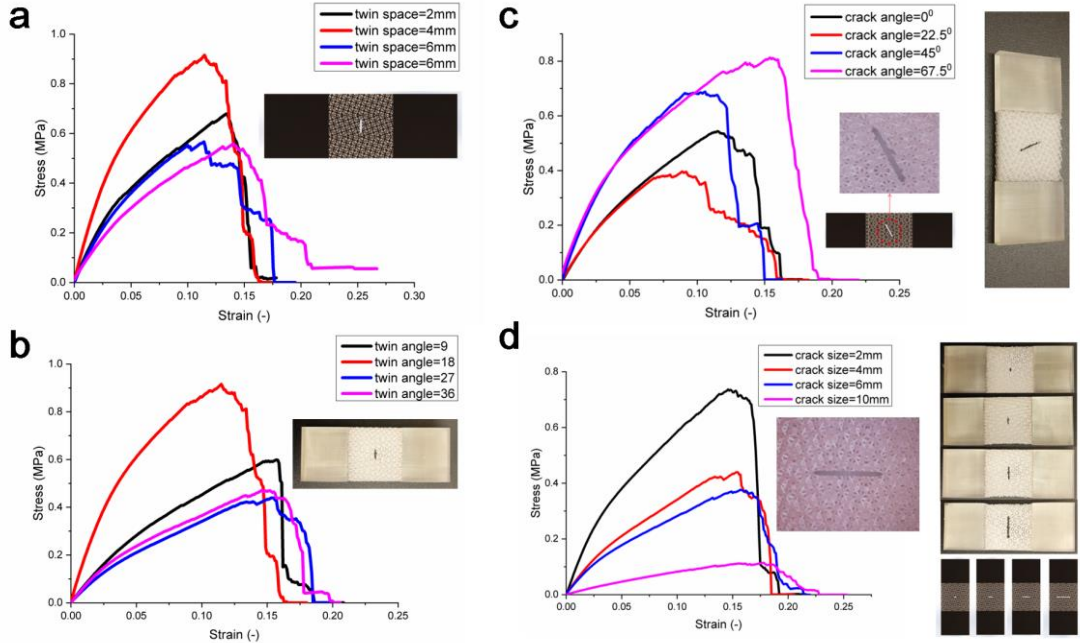

**Fig. 6: Uniaxial tensile strain–stress curves of TMMs containing centred cracks.** (a) Effects of twin-space width on crack-propagation resistances, where the sample possesses a twin-space width of *t* = 4 mm, exhibiting strongest crack resistance. (b) Effects of twin angle on crack-propagation resistances; twin angle of sample is **θ**= 18°and exhibits strongest crack resistance. (c) Strain–stress curves of tensile samples containing centred cracks with different spatial angles; the sample with spatial angle of **β** = 22.5°exhibits weakest crack resistances. (d) Uniaxial tests of TMMs with pre-existing cracks for studying crack-size effects on crack-propagation resistance.

**Conclusions**

In this study, we proposed crystal-inspired TMMs and demonstrated their unique mechanical performances. Our findings revealed that TMMs can improve the energy-absorption efficiency of the stretching-dominant lattice structure (from approximately 59% to 96%), and can maintain the plateau stress of standard FCC mechanical metamaterials (without twinning). In addition, TMMs can improve the anti-fracture toughness of standard FCC mechanical metamaterials. Furthermore, we reported on the inverse Hall–Petch effect of TMMs. in the case of graded TMMs, the crack propagation resistances from coarse twin to fine twin of TMMs is analogous to the brittle-ductile transition behaviours of graded twin nano-crystals. We also systematically investigated the effects of twin angle, twin-space, crack size, and crack spatial orientation on crack-propagation resistances. Our findings could help in the designing and fabrication of advanced materials with exceptional tuneable mechanical properties.

**Methods**

**Design:** Based on the structural twin features in nano-crystals and biomaterials, we propose novel ordinary twin, graded and hierarchical TMMs for energy absorption. First, periodic architected FCC mechanical metamaterials are rotated at a certain angle, and the corresponding twin layers are cut out. Then, mirror operations are performed to harvest desirable TMM samples. The rotated angle related to the twin interface is called the twin angle ($\theta$) and the distance between neighbouring twin interfaces is called twin-space ($d$), as shown in **Fig. 1(d)**. Similarly, as shown in **Fig. 1(e)**, a graded design can be realised by adjusting the geometrical relation between neighbouring twin layers, for example, twin angle $\theta$ or twin-space. Moreover, as shown in **Fig. 1(f)**, hierarchical TMMs can be designed through repeated twinning operations at different structural scales, where the twin-space widths at different structural levels are denoted as $t_1$ and $t_2$, and the twin angles at the corresponding levels are denoted as $\theta_1$ and $\theta_2$, respectively.

**Materials and fabrication:** In this work, we used the Ember 3D printer (Autodesk, USA), which is based on the PμSL technique using a photocurable polymer, to fabricate TMMs. The Autodesk standard clear prototyping resin (PR48) from Colorado Photopolymer Solutions was employed as the material for printing. The TMMs of the designed file were sliced to a thickness of 50 μm using the print studio software (Autodesk, USA). The exposure time was adjusted from 7 to 20 s for each sliced layer by considering the structural complexity. After 3D printing from the resin pool, the printed samples were cleaned with isopropyl alcohol (IPA) for 10 min to dissolve the residual liquid resin from the samples. Then, the samples were solidified by post-curing through UV illumination for 30 min. The tested mechanical properties of the material for 3D printing are shown in **supplementary Table S2**. The process of 3D-printing-based fabrication is shown in **supplementary Fig. S1**, and examples of as-fabricated ordinary, graded, and hierarchical TMMs are shown in **Figs. 1(g)**, **1(h)**, and **1(i)**, respectively.

**Mechanical tests and analyses**

*Compression test*: In-situ compression tests on an Instron@ 3400 machine were performed for studying the energy-absorption efficiency of TMMs, where the loading cell range reaches 2000N, with force and displacement resolutions of 0.01N and 0.002mm, respectively. In situ compression with a loading speed of 0.2mm/min was applied in all compression tests, and images of the deformation process was recorded at a recording speed of 1 pic/s.

*Tensile test*: Tensile tests were performed on the Instron@ 3400 machine to study the relation among the tensile strength, crack-propagation resistance, crack size, crack orientation, twin size, and twin-space of TMMs; the loading cell range is 2000N, with force and displacement resolutions of 0.01N and 0.002mm, respectively. In situ compression with a loading speed of 0.2mm/min was applied for all tensile tests, and images of the deformation process were recorded at a recording speed of 1 pic/s.

*Three-point bending test*: Three-point bending tests were performed on the Instron@5900 machine to understand the mechanical benefits of twinning on the crack-propagation resistances of TMMs. The fracture crack resistances of FCC unit-cell-based mechanical metamaterials are tested in mode I by using a singled-edge notch 3-point bending sample, following the procedure outlined in ASTM standard E1820 (ASTM E1820, 2013). Load spreaders placed between the rollers and sample prevented local indentation, and the two outer rollers define the span [46]. The loading cell range of the 3-point bending test machine was 100N, with force and displacement resolutions of 0.001N and 0.005mm. Insitu compressions with the loading speeds of 0.2mm/min were applied in all the 3-point bending tests, and images of the deformation process were recorded at a recording speed of 1 pic/s.

**Acknowledgments**

SK acknowledges the National Research Foundation of Korea (NRF) grant funded by the Korea government (MSIT) (NRF-2019R1A5A808320112). YTC acknowledge the support of the Technology Innovation Program (20007064, Realization of air cleaning mobility HAMA (superHydrophobic Additive Manufactured Air cleaner) Project funded by the Ministry of Trade, Industry, & Energy (MOTIE, Korea). WW acknowledges National Natural Science Foundation of China (Grant No. 11702023; No. 11972081) is acknowledged.

# Supplementary Information

Wenwang Wu[1,2*#], Seok Kim[1,3*], Ali Ramazani[1*], Young Tae Cho[3], Daining Fang[4#]

1, Department of Mechanical Engineering, Massachusetts Institute of Technology, Cambridge, MA 02139, USA.

2, School of Naval Architecture, Ocean and Civil Engineering, Shanghai Jiao Tong University, Shanghai, 200240, China.

3, Department of Mechanical Engineering, Changwon National University, Changwon, South Korea.

4, Institute of Advanced Structures & Technologies, Beijing Institute of Technology, Beijing, 100084, China.

\* equally contribution

\# corresponding authors

Wenwang Wu    (wuwenwang@sjtu.edu.cn)

Daining Fang    (fangdn@pku.edu.cn)

**Table. S1: Sample terminology lists.**

Geometrical design of compression samples and uniaxial compression tests are performed along the *z* direction. The geometrical details of seven samples (standard without twin, ordinary twin, graded twin, and hierarchical twin) are shown below, where the FCC unit cell size is *s*=3.2mm and the constituent beam diameter is *d*=0.4mm.

| No. | Twin design categories | Details |
|---|---|---|
| A | Standard FCC mechanical metamaterials as reference sample for comparison. | standard FCC unit cell periodically architected mechanical metamaterials, sample sizes are 32, 25.6, 25.6mm, consisting of $10 \times 8 \times 8$ unit cells along the *x*, *y*, and *z* directions, respectively. |
| B | Ordinary twin design with identical twin-spaces along the *z* direction. | twin-space width, $t$ =6.4mm; twin angle, $\theta = 30°$; sample sizes along x, y and z directions: 32, 25.6, and 25.6mm, with 4 layers of twin designs along the *z* direction. |
| C | Ordinary twin design with identical twin-space along the *z* direction. | twin-space width, $t$ =6.4mm; twin angle, $\theta = 15°$; sample sizes along x, y and z directions 32, 25.6, and 25.6mm, comprising 4 layers of twin designs along the *z* direction, respectively. |
| D | Graded twin design consisting of 3 layer thicknesses, with increasing twin-space from bottom to top along the *z* direction. | designed with increasing layer thickness from bottom to top along the *z* direction; the thickness values of the 3 twin layers are $t_1$=3.2mm, $t_2$=6.4mm, and $t_3$=12.8mm. The TMM sample is constructed with $4t_1$, $2t_2$, and $1t_3$ layers. The twin angle is $\theta = 30°$ and sample sizes are 32mm, 25.6mm and 38.4mm along the *x*, *y* and *z* directions, respectively. |
| E | Graded twin design consisting of 3 layer thicknesses, with increasing twin-space from bottom to top along the *z* direction. | designed with increasing layer thickness from bottom to top along the *z* direction; the thicknesses values of the 3 twin layers are $t_1$=3.2mm, $t_2$=6.4mm, and $t_3$=12.8mm. The TMM sample is constructed with $4t_1$, $2t_2$, and $1t_3$ layers. The twin angle is $\theta = 15°$ and sample sizes are 32mm, 25.6mm and 38.4mm along the *x*, *y* and *z* directions, respectively. |
| F | Hierarchical TMMs consisting of primary and secondary twin topological features. | The primary twin-space is $t_P$=12.8mm, with the twin angle of $\theta_P = 5°$. The secondary twin-space is $t_S$=6.4mm, with the secondary twin angle of $\theta_S = 25°$. The sample sizes are 51.2mm, 51.2mm and 38.4mm along the *x*, *y* and *z* directions, respectively. Accordingly, 4 and 8 layers of primary and secondary twin designs exist along the *z* and *x* directions, respectively. |
| G | Hierarchical TMMs consisting of primary and secondary twin topological features. | The primary twin-space is $t_P$=12.8mm, with the twin angle of $\theta_P$=20°. The secondary twin-space is $t_S$=6.4mm, with the secondary twin angle of $\theta_S = 25°$. The sample sizes are 51.2mm, 51.2mm and 38.4mm along the *x*, *y* and *z* directions, respectively. Accordingly, 4 and 8 layers of primary and secondary twin designs exist along the *z* and *x* directions, respectively. |

**Table. S2: Mechanical properties of the materials used for 3D printing [Ref S1].**

|  | Fresh print | Post-cured |
|---|---|---|
| Tensile modulus (MPa) | 600 | 1400 |
| Tensile strength (MPa) | 16 | 28 |
| Elongation (%) | 5 | 3 |

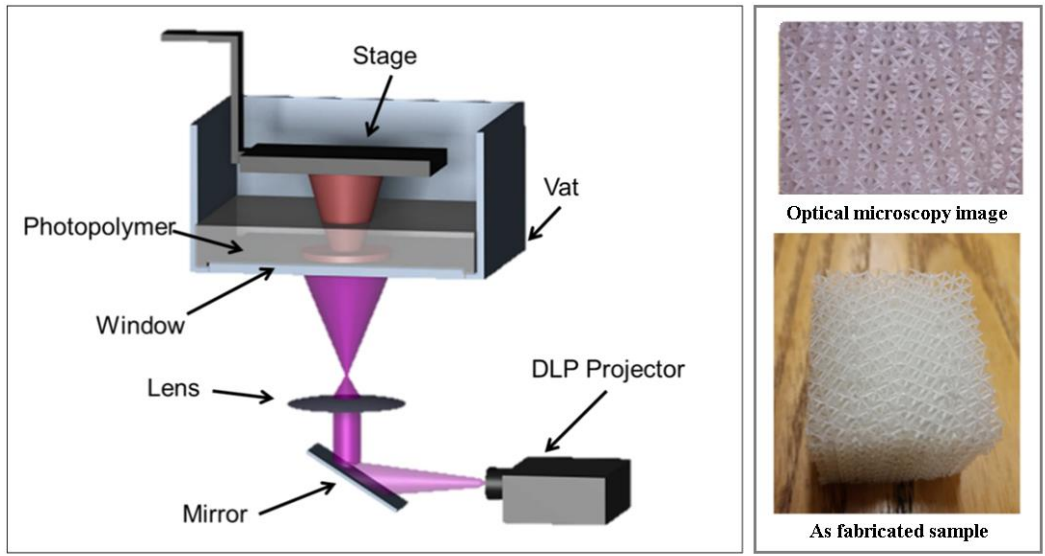

**Fig. S1**: Projection microstereolithography (PμSL) printing for fabrication of twin mechanical metamaterials (TMMs).

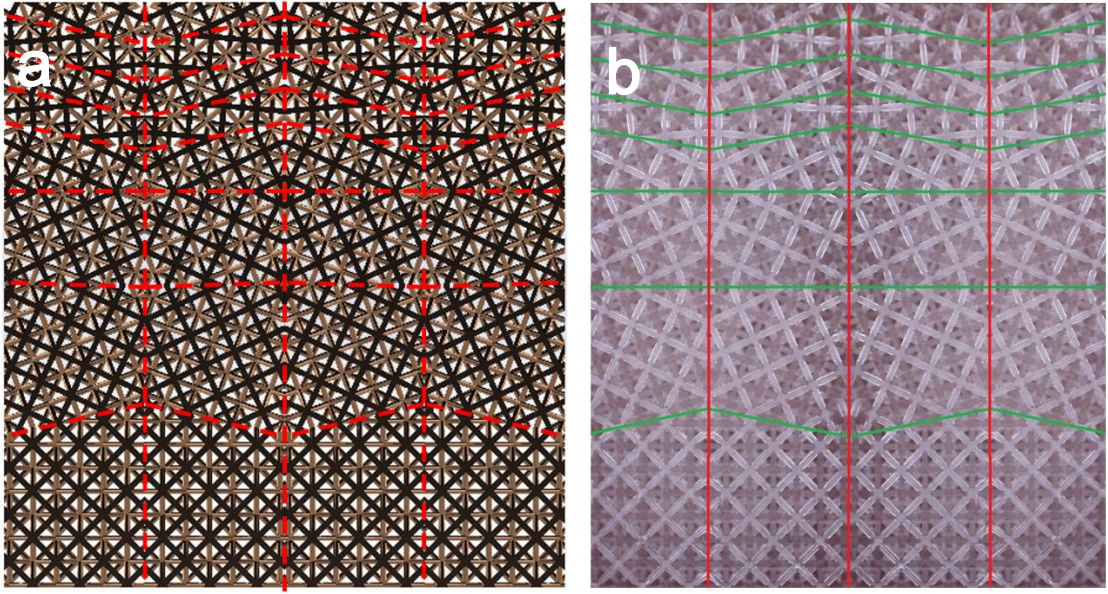

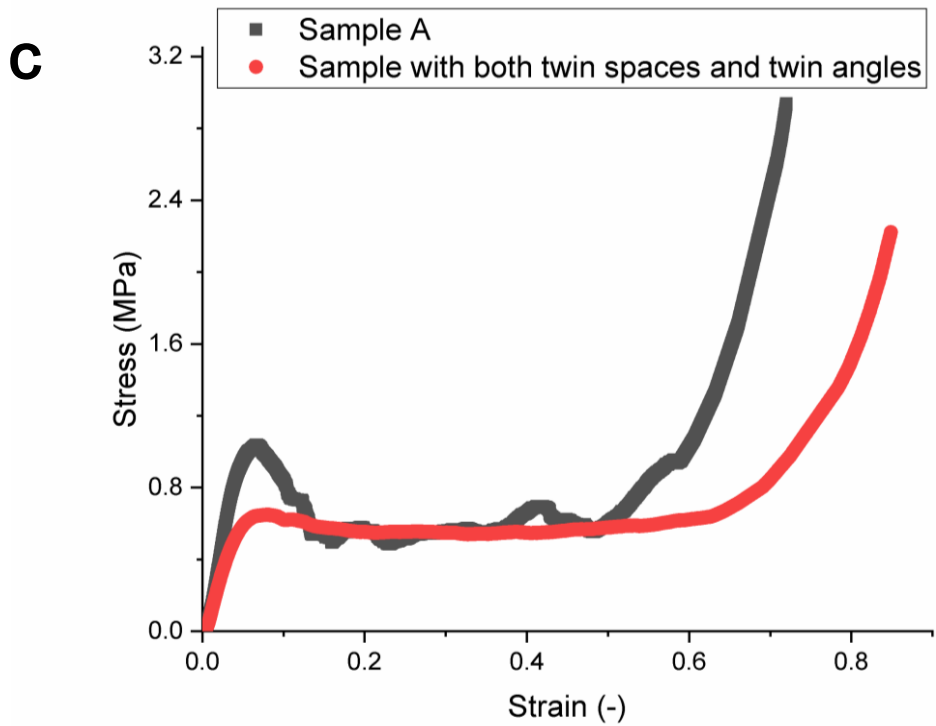

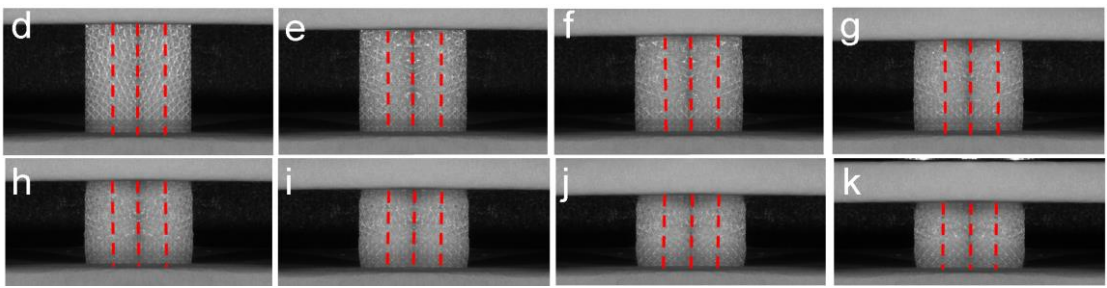

**Fig. S2:** (a) as designed gradient twin sample with gradient angle (twin angle = 0°, 20° and 30°) and gradient spaces (2.4mm, 4.8mm and 7.2mm), where the FCC unit cell size is $s$=3.2mm and the constituent beam diameter is $d$=0.4mm, and sample sizes are: 38.4mm, 40mm and 19.2mm

along the *x*, *y* and *z* directions, respectively; (b) as fabricated gradient twin sample with gradient angle (twin angle = 0°, 20° and 30°) and gradient spaces (2.4mm, 4.8mm and 7.2mm); (c) compression strain-stress curves of TMMs with both graded twin spaces and twin angles, the resultant peak stress, mean plateau stress and energy absorption efficiency are: 0.6493MPa, 0.6111MPa and $\varphi$=0.94, respectively. Meanwhile, the resultant peak stress, mean plateau stress and energy absorption efficiency of standard periodic architected FCC sample without twin topology design are: 1.0304MPa, 0.6099MPa and $\varphi$=0.59, respectively; (d)-(k) compression deformation process at different strain level ε=0.00, 0.0576, 0.1126, 0.1764, 0.2278, 0.2855, 0.3413, 0.4016, respectively.

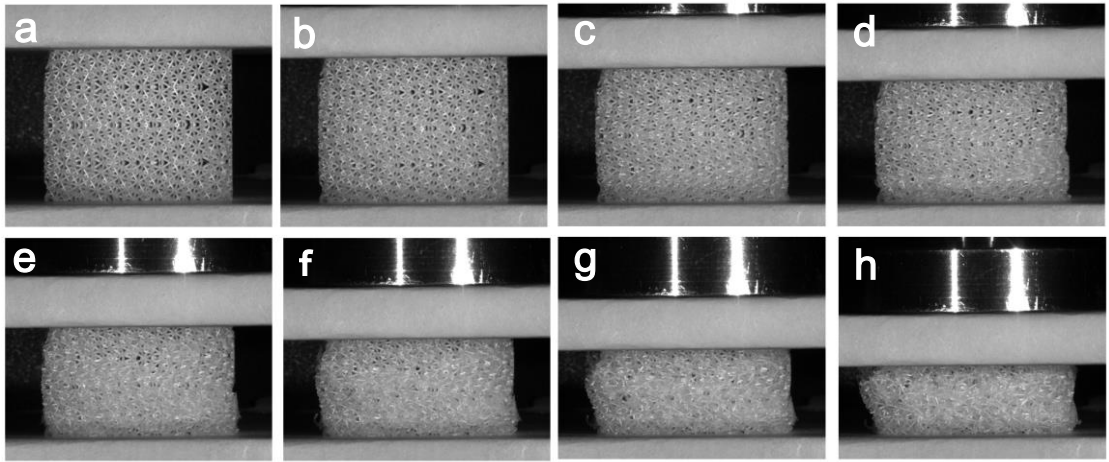

**Fig. S3:** Local deformation and failure process of the TMMs (sample **B**) before densification during the compression process, showing local deformation features at different strain levels ε of (a) 0.0, (b) 0.0539, (c) 0.1320, (d) 0.2101, (e) 0.2896, (f) 0.3676, (g) 0.4464, and (h) 0.5550, respectively.

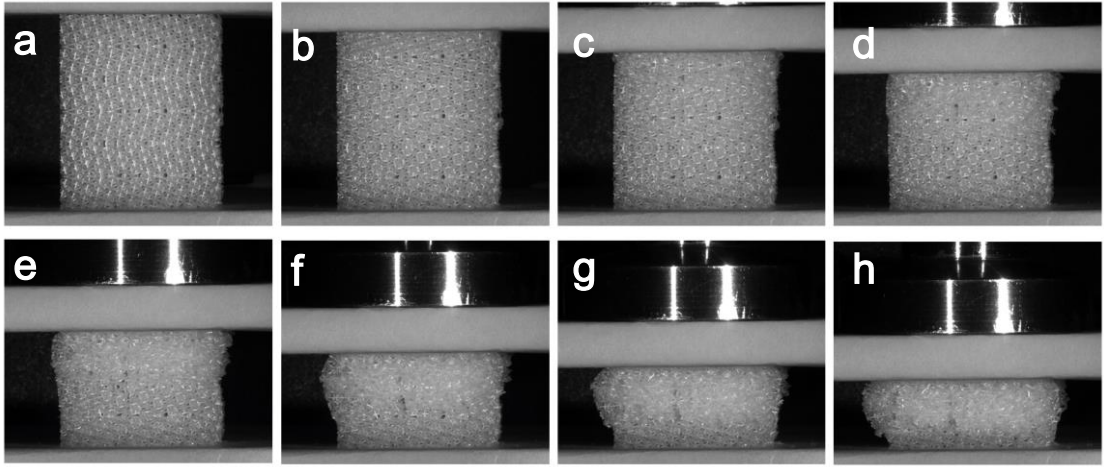

**Fig. S4:** Compression process and microstructure evolution process of graded TMMs (sample **D**), showing local deformation features at different strain levels ε of (a) 0.0, (b) 0.0901, (c) 0.1970, (d) 0.3039, (e) 0.4107, (f) 0.5176, (g) 0.5887, and (h) 0.6525, respectively.

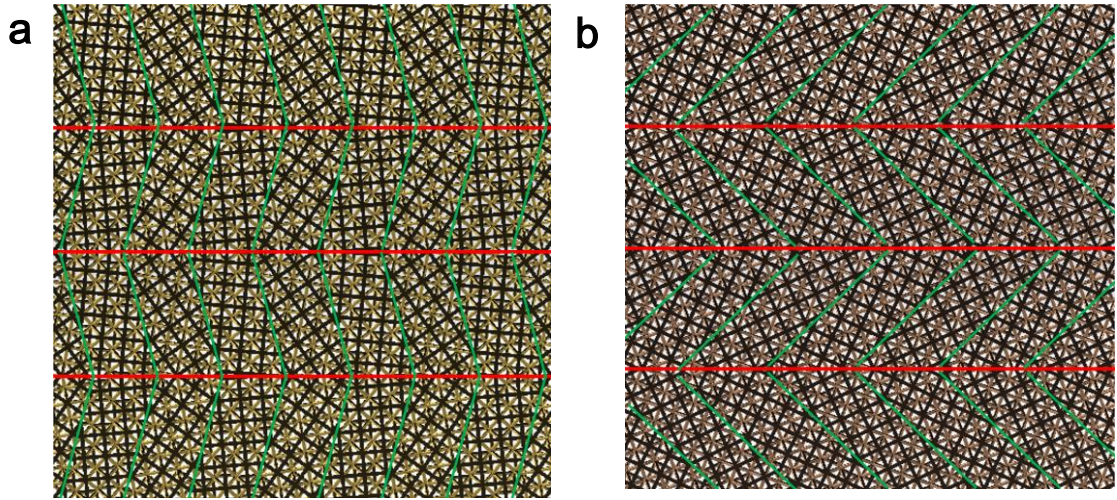

**Fig. S5**: Microstructures of hierarchical TMMs: (a), sample **F,** and (b), sample **G**.

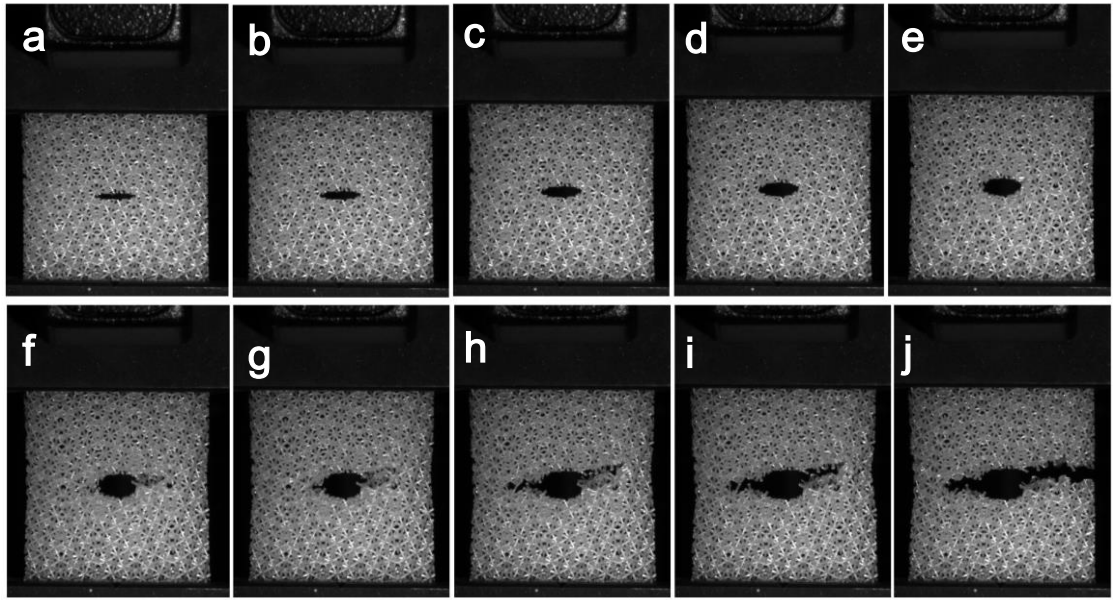

**Fig. S6:** Microstructure evolution of TMMs for studying effects of twin-space width and twin angle on crack-propagation resistances, where crack propagation features of a typical tensile sample (designed with twin-space of 4 mm and twin angle of $\theta = 18°$) containing a centred crack are shown at strain level of (a) 0.0, (b) 0.0289, (c) 0.0553, (d) 0.0825, (e) 0.1105, (f) 0.1369, (g) 0.1437, (h) 0.1497, (i) 0.1565, and (j) 0.1607, respectively.

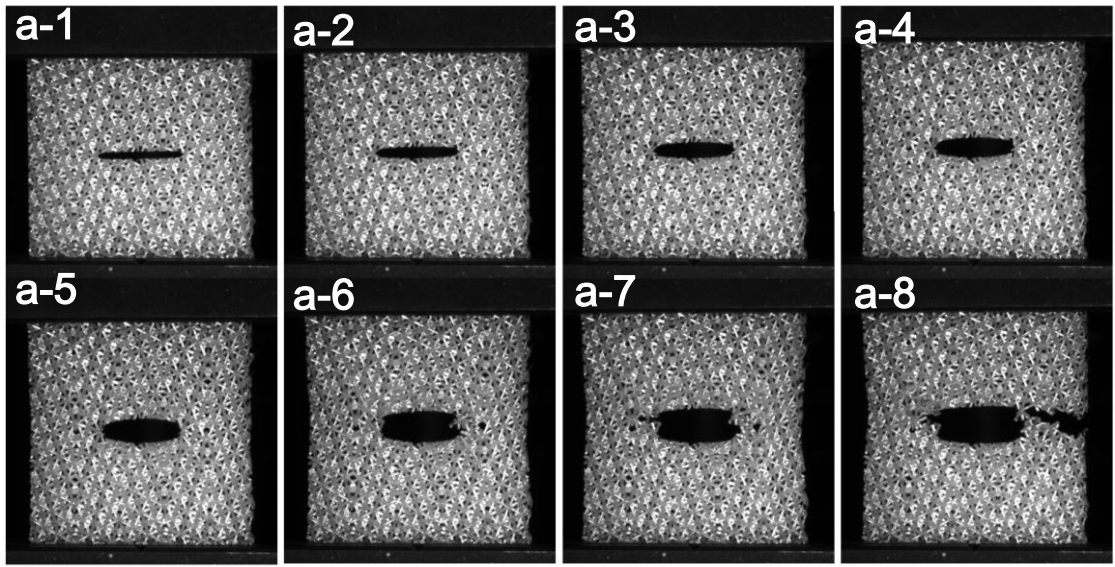
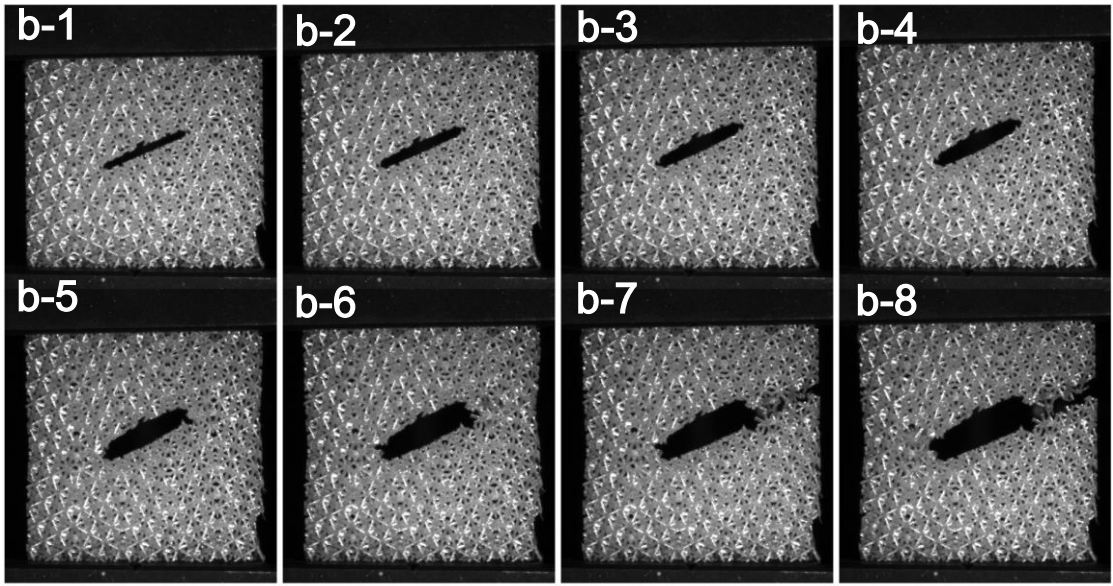
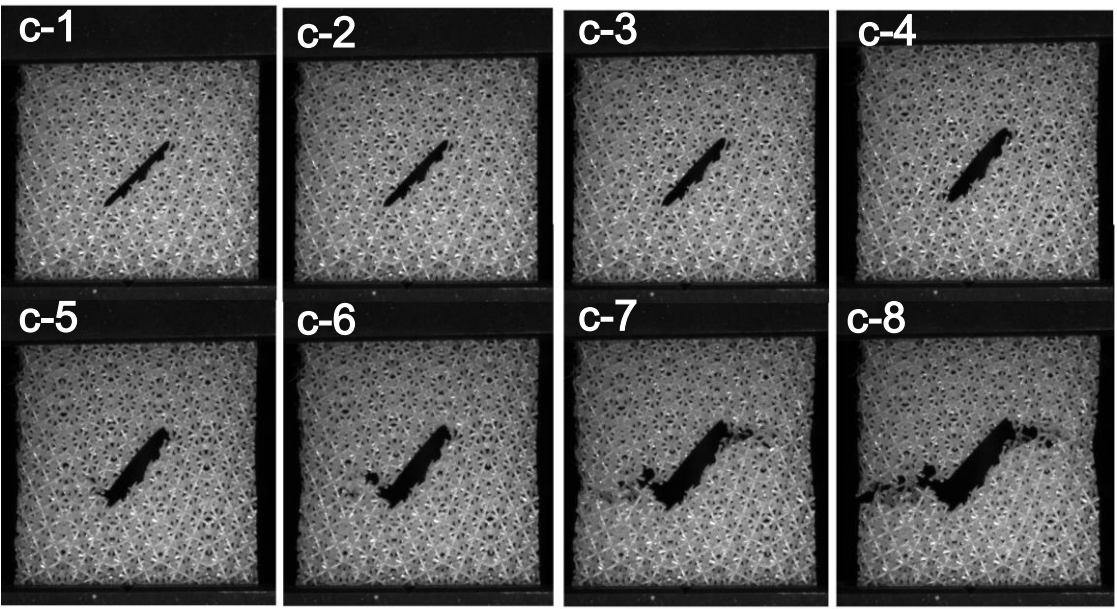

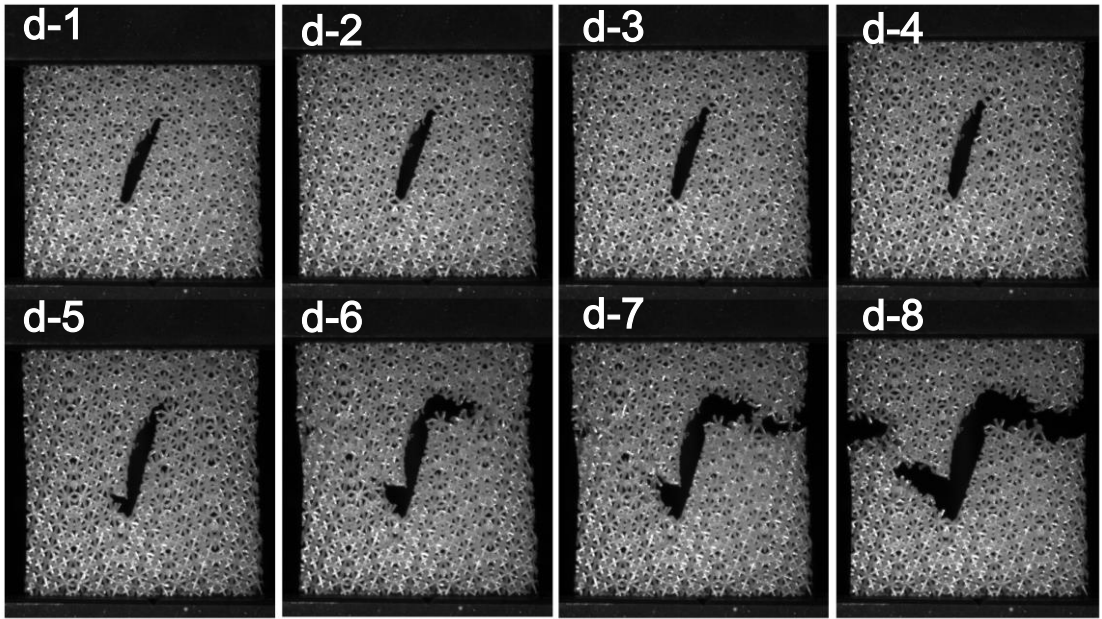

**Fig. S7**: Microstructure evolution of TMMs for studying effects of crack-orientation angle on crack-propagation resistances, where crack propagation features with different crack spatial orientations within the middle of the TMMs are shown at different strain levels of (a-1) 0.0, (a-2) 0.0286, (a-3) 0.0556, (a-4) 0.0825, (a-5) 0.1111, (a-6) 0.1380, (a-7) 0.1498, and (a-8) 0.1549 respectively; (b-1) 0.0, (b-2) 0.0203, (b-3) 0.0439, (b-4) 0.0650, (b-5) 0.0861, (b-6) 0.1097, (b-7) 0.1156, and (b-8) 0.1300 respectively; (c-1) 0.0, (c-2) 0.0220, (c-3) 0.0423, (c-4) 0.0880, (c-5) 0.1100, (c-6) 0.1252, (c-7) 0.1345, and (c-8) 0.1438 respectively; (d-1) 0.0, (d-2) 0.0354, (d-3) 0.0673, (d-4) 0.1110, (d-5) 0.1564, (d-6) 0.1775, (d-7) 0.1842, and (d-8) 0.2052 respectively.

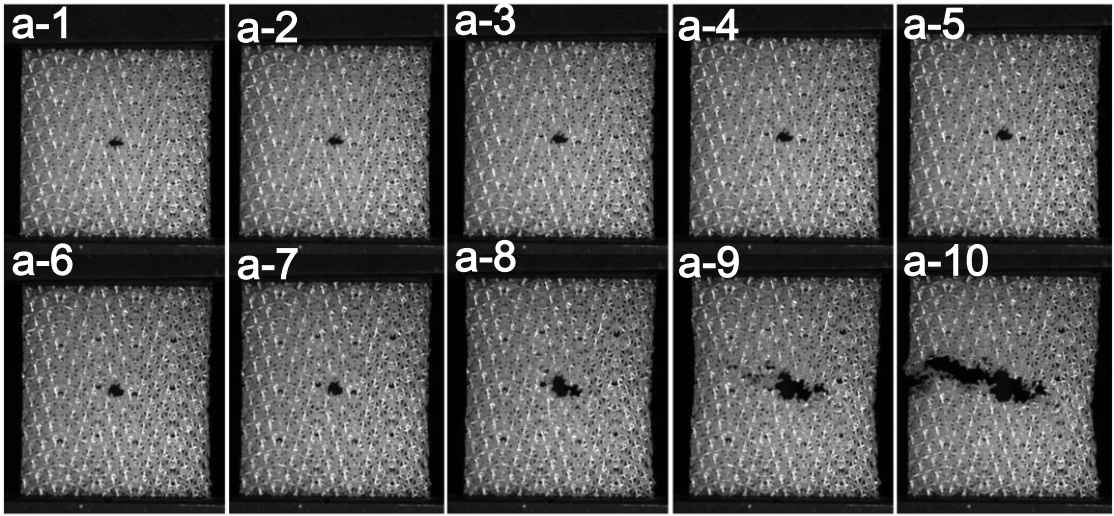
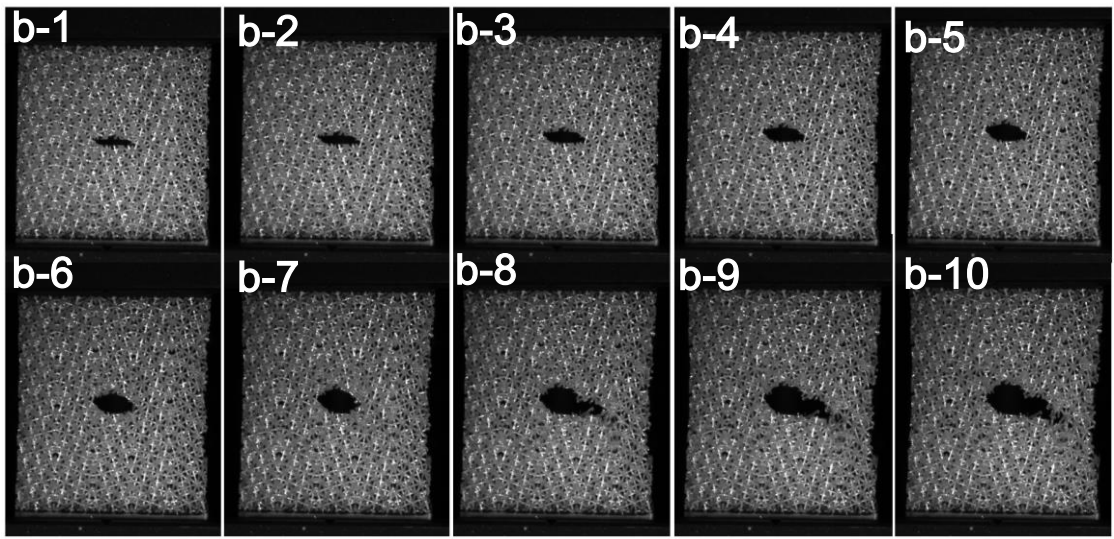
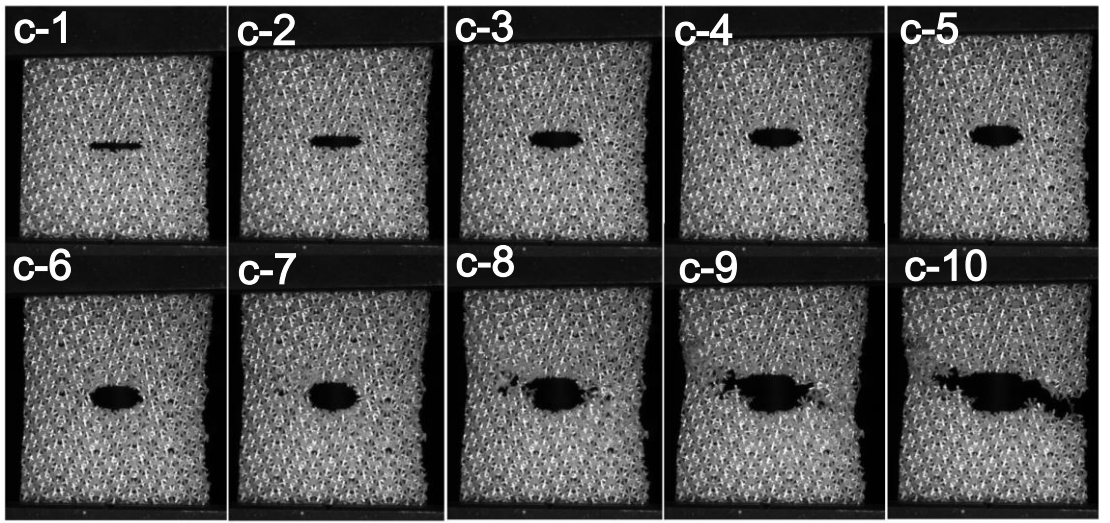

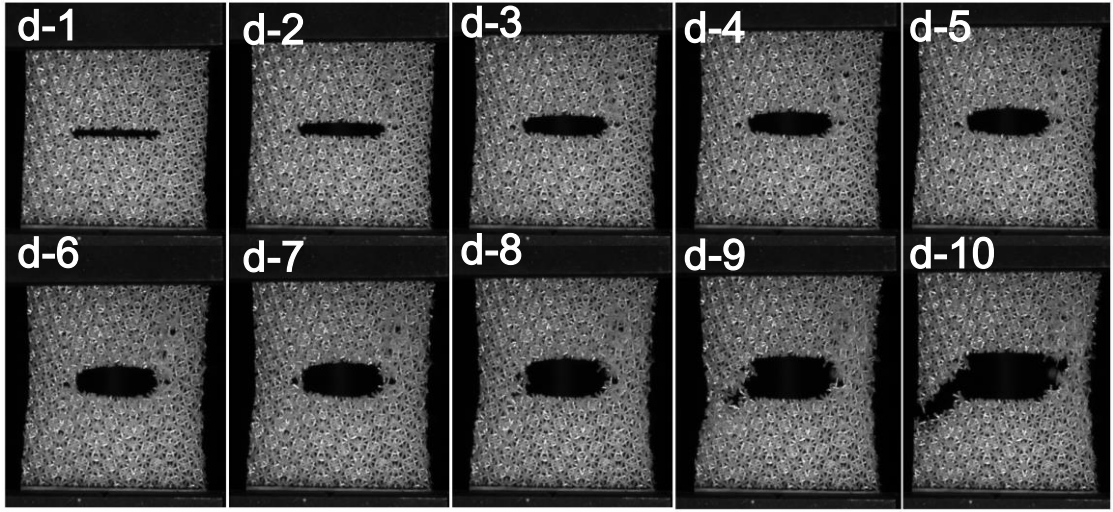

**Fig. S8**: Microstructure evolution of TMMs for studying crack size effects on crack-propagation resistances, where crack-development features during deformation process are shown at different strain levels of (a-1) 0.0, (a-2)0.01891, (a-3)0.0393, (a-4)0.0590, (a-5)0.0794, (a-6)0.1172, (a-7)0.1362, (a-8)0.1566, (a-9)0.1619, and (a-10)0.1672, respectively; (b-1)0.0, (b-2)0.0256, (b-3)0.0468, (b-4)0.0717, (b-5)0.0958, (b-6)0.1192, (b-7)0.1419, (b-8)0.1485, (b-9)0.1529, and (b-10)0.1580, respectively; (c-1) 0.0407, (c-2) 0.0844, (c-3) 0.1024, (c-4) 0.1251, (c-5) 0.1454, (c-6) 0.1658, (c-8) 0.1759, (c-9) 0.1869, and (c-10)0.1923, respectively; (d-1) 0.0, (d-2) 0.0422, (d-3) 0.0820, (d-4) 0.1038, (d-5) 0.1233, (d-6) 0.1436, (d-7) 0.1663, (d-8) 0.1850, (d-9) 0.2077, and (d-10)0.2287, respectively.

**References:**

[S1] https://cpspolymers.com/PR48%20TDS.pdf